\documentclass[a4paper, twocolumn]{extarticle}
\usepackage[utf8]{inputenc}
\usepackage[T1]{fontenc}

\usepackage{geometry}
\usepackage{fancyhdr}
\usepackage{graphicx}

\usepackage{hyperref}
\hypersetup{colorlinks=true,linkcolor=blue,urlcolor=blue,citecolor=blue}

\usepackage{amsmath,amssymb,mathrsfs}

\numberwithin{equation}{section}

\usepackage{kpfonts}

\usepackage[affil-it]{authblk}

\usepackage{abstract}

\usepackage{titlesec}
\titleformat*{\section}{\large\bfseries}

\usepackage[labelfont={bf},textfont={small},width=230pt]{caption}

\usepackage{cite}
\title{ \huge An illustration of canonical \\ quantum-classical dynamics \\
    \vspace{0.2cm}
    \Large {Backreaction, canonical relations and time evolution in the quantum-classical harmonic oscillator}
}

\author[1]{Mustafa Amin \thanks{\href{mailto:m.amin@uleth.ca}{m.amin@uleth.ca}}}
\author[2]{Mark A.~Walton \thanks{\href{mailto:walton@uleth.ca}{walton@uleth.ca}}}

\affil[1,2]{ 
    Department of Physics \& Astronomy, University of Lethbridge, Lethbridge, Alberta, T1K 3M4, Canada
}

\date{}

\begin{document}

\twocolumn[
    \begin{@twocolumnfalse}

\maketitle

\begin{abstract}
    Using the example of the harmonic oscillator, we illustrate the use of hybrid dynamical brackets in analyzing quantum-classical interaction.  We only assume that a hybrid dynamical bracket exists, is bilinear, and reduces to the pure quantum/classical bracket when acting on pure quantum/classical variables.  Any hybrid bracket obeying these natural requirements will produce the same dynamics for pure classical or quantum variables, given a hybrid Hamiltonian.  Backreaction is manifested in the evolution of a nonvanishing commutator between classical variables.  The more massive the classical system is, the less it is affected by backreaction.  Interestingly, we show that while pure variables evolve to violate the pure canonical relations, they always obey the hybrid canonical relations.  The dynamics of hybrid variables, on the other hand, is shown to require a fully specified and consistent hybrid bracket, otherwise evolution cannot be defined.

\vspace{1cm}

\end{abstract}

    \end{@twocolumnfalse}
]

\saythanks

\section{Introduction}
\label{intro}

Quantum-classical systems appear in numerous fields, from chemical and condensed matter physics to cosmology and black hole physics.  In practical applications, a full quantum treatment is often computationally intractable.  Treating parts of the system as classical simplifies calculations and provide easier models to conceptualize and approach such problems.  We constantly deal with quantum electrons around classical nuclei, quantum matter on classical spacetimes or quantum systems interacting with external classical fields.  Less practically, foundational research on the nature of gravity or the measurement problem in quantum mechanics may also benefit from the study of quantum-classical systems.

Familiar approaches to modelling quantum-classical interactions usually fall into two categories.  The first treats the classical part of the system as omnipotent; it acts on, but is not acted back on by, the quantum part.  Examples of this type include textbook quantum mechanics~\cite{maddox_classical_1995}, or quantum field theories on curved spacetime.

The second approach is to use averages of quantum observables $\langle x_Q \rangle$ in classical equations
$$\text{Eq.}_\text{Classical} \big(\langle x_Q \rangle \, , y_C \big)~,$$
and then feed the classical observables $y_C$ back into quantum equations
$$\text{Eq.}_\text{Quantum} \big(x_Q \, , y_C \big)~.$$
While this approach captures a form of quantum backreaction on the classical system, it suffers from two problems: 1.~it averages out quantum fluctuations so they do not manifest on the classical side, and 2.~it introduces nonlinearity to the quantum Schrödinger equation thus negating the superposition principle.  Even if one ignores the loss of superposition as a side effect of a simplifying approximation, nonlinear equations can be very difficult to handle in practical applications.

In this paper, we are concerned with a third approach: directly coupling quantum and classical variables.  This requires defining a consistent framework for the dynamics of such hybrid quantum-classical variables.  Namely, we are interested in a canonical dynamical structure where the equations of motion are defined in terms of a hybrid dynamical bracket and canonical relations between conjugate variables.  Many proposals for a hybrid quantum-classical bracket can be found in the literature~\cite{aleksandrov_statistical_1981,gerasimenko_dynamical_1982,boucher_semiclassical_1988,anderson_quantum_1995,prezhdo_mixing_1997,prezhdo_quantum-classical_2006,elze_linear_2012,bondar_koopman_2019}.  See also~\cite{barcelo_hybrid_2012,elze_four_2012,elze_quantum-classical_2013}.

However, impediments to a consistent hybrid dynamics have been found~\cite{salcedo_absence_1996,caro_impediments_1999,sahoo_mixing_2004,salcedo_statistical_2012,gil_canonical_2017}.  Specifically, it was shown that hybrid dynamical brackets do not satisfy the Jacobi identity or the Leibniz rule.  These are crucial properties for dynamical consistency.  The Jacobi identity guarantees that the fundamental canonical relations are preserved with dynamical evolution.  The Leibniz rule is essential for defining dynamical evolution and derivatives, as shall be demonstrated in Sec.~\ref{sec:4}.

The present authors, however, have shown in~\cite{amin_quantum-classical_2020} that starting from a full quantum system and taking the classical limit of a part, one can uncover a general class of hybrid brackets.  Different brackets arise from different quantization schemes on the classical sector prior to taking the classical limit.  In particular, the bracket proposed by Aleksandrov~\cite{aleksandrov_statistical_1981}, Gerasimenko~\cite{gerasimenko_dynamical_1982}, and Boucher and Traschen~\cite{boucher_semiclassical_1988} arises as a special case of the general bracket when Wigner-Weyl quantization is used.  The derivation of the general bracket makes use of the phase space formulation of quantum mechanics showing a connection between phase space distributions (like that of Wigner, Husimi, etc.)  and hybrid dynamical brackets.

To address the aforementioned no-go theorems, the authors introduced a \textit{hybrid composition product}~\cite{amin_quantum-classical_2020}.  The general hybrid bracket is then the commutator of that product.  It follows that the Jacobi identity and Leibniz rule are automatically satisfied provided that the hybrid composition product is associative.

Then, for a restricted set of hybrid variables that form an associative subalgebra with the hybrid composition product, dynamics is consistent.  Since quantum-classical interactions are represented by hybrid terms in the Hamiltonian, the consistency of quantum-classical dynamics in our scenario implies that only certain interactions between quantum and classical systems are allowed.

Here we present a working example of canonical hybrid dynamics that provides a blueprint for applying the framework to possible problems of interest.  The harmonic oscillator is ubiquitous in physical systems.  We use it to illustrate the features of hybrid dynamics and the cautionary pitfalls of dealing with systems that are not clearly classical or quantum.

Notably, hybrid evolution of pure variables does not depend on any particular definition of a hybrid bracket as long as it is bilinear and reduces to a pure bracket when one of its arguments is pure.  Evolution of hybrid variables, on the other hand, requires specification of the hybrid bracket used.  We emphasize that the methods presented in this paper are applicable to more general problems.

We introduce basic concepts and general properties of hybrid brackets in Sec.~\ref{sec:2}.  In Sec.~\ref{sec:3}, we calculate the equations of motion of pure classical and quantum variables in the quantum-classical harmonic oscillator.  Adopting Anderson's argument~\cite{anderson_quantum_1995}, we find the ``backreaction'' of the quantum on the classical and show the blurring of the line between quantum and classical variables due to interaction.

Finally, we describe an example of time evolution of hybrid variables in Sec.~\ref{sec:4}.  In that example, the need for a consistent hybrid bracket is illustrated.  Different brackets, corresponding to different quantization schemes (and, by extension, different phase space distributions) are shown to produce different time evolutions for hybrid variables.

\section{Hybrid dynamics}
\label{sec:2}

There are multiple approaches to the problem of combining quantum and classical systems~\cite{barcelo_hybrid_2012,elze_four_2012,elze_quantum-classical_2013}.  In this paper, we are concerned with the canonical approach, in which the problem is that of defining a hybrid dynamical bracket.  The hybrid bracket $\{\![\cdot\,,\cdot]\!\}$ then plays, for hybrid variables, the role of the commutator $[\cdot\,,\cdot]/i\hbar$ for quantum variables or the Poisson bracket $\{\cdot\,,\cdot\}$ for classical ones.

Hybrid brackets have been proposed in~\cite{aleksandrov_statistical_1981,gerasimenko_dynamical_1982,boucher_semiclassical_1988,anderson_quantum_1995,prezhdo_mixing_1997,prezhdo_quantum-classical_2006,elze_linear_2012,bondar_koopman_2019}.  The authors have derived a general class of hybrid brackets in~\cite{amin_quantum-classical_2020} by applying a partial classical limit to quantum mechanics.  Different arguments may produce different brackets, but a general feature of all of them is their reduction property.  A hybrid bracket should reduce to a pure (quantum or classical) bracket when one of its arguments is pure:
\begin{align}\label{eq:reduction}
\begin{split}
    \{\![\cdot\,,\,A_Q]\!\} &= \frac{1}{i\hbar} [\cdot\,,\,A_Q]~, \\
    \{\![\cdot\,,\,A_C]\!\} &= \{\cdot\,,\,A_C\}~.
\end{split}
\end{align}
The subscripts $Q$ and $C$ signify quantum and classical variables, respectively.  Another property shared by hybrid brackets is bilinearity:
\begin{align}\label{eq:lin}
\begin{split}
    \{\![A, B + C]\!\} = \{\![A, B]\!\} + \{\![A, C]\!\}~, \\
    \{\![A + B, C]\!\} = \{\![A, C]\!\} + \{\![B, C]\!\}~.
\end{split}
\end{align}

A general Hamiltonian is given by
\begin{align}\label{eq:hamiltonian}
    H = H_Q + H_C + V~.
\end{align}
While the pure $H_Q$ and $H_C$ encode internal dynamics of the quantum and classical subsystems, $V$ is a hybrid interaction term that couples quantum and classical variables.  Without knowing the specific definition of the hybrid bracket, one can find the equation of motion for pure variables using only the reduction requirement~\eqref{eq:reduction}
\begin{align}\label{eq:eom-pure}
    \frac{d}{dt}A_\text{Pure} = \{\![A_\text{Pure},H]\!\} + \frac{\partial}{\partial t} A_\text{Pure}~.
\end{align}
The equation is given in the Heisenberg picture, where time evolution resides in the dynamical variables instead of the state.

The reduction requirement~\eqref{eq:reduction} can go a long way in defining hybrid dynamics as shown by an explicit example in Sec.~\ref{sec:3}.  Specifically, when one is interested in calculating the time evolution of pure variables, as in Eq.~\eqref{eq:eom-pure}, one can rely on the well-defined quantum or classical brackets to perform calculations.

The dynamics of hybrid variables is more complicated, however.  It was shown in~\cite{salcedo_absence_1996,caro_impediments_1999,sahoo_mixing_2004,salcedo_statistical_2012,gil_canonical_2017} that hybrid brackets cannot define a consistent dynamical framework for general hybrid variables.

To deal with the problem of evolving hybrid variables or, equivalently, having both arguments of the dynamical bracket be hybrid, we need a specific definition of the bracket.  The authors have derived a general hybrid bracket in~\cite{amin_quantum-classical_2020} along with a definite consistency condition for hybrid brackets.  Namely, only a certain subset of hybrid variables is allowed into the theory for which hybrid dynamics is consistently defined.  An immediate consequence of such restriction is that the consistency of hybrid dynamics dictates what kind of interaction could exist between quantum and classical systems.  The derivation of the general hybrid bracket in~\cite{amin_quantum-classical_2020}, as outlined in Sec.~\ref{sec:4}, relies on the methods of phase space quantum mechanics.

\section{Backreaction in the harmonic oscillator}
\label{sec:3}

One of the main motivations for developing hybrid dynamics is to study the effect of quantum systems on classical ones.  It is to be expected that initially deterministic classical variables will inherit some uncertainty by interacting with a quantum system.  In~\cite{anderson_quantum_1995}, Anderson argues that classical variables will evolve to become ``quasiclassical'': variables that exhibit only \textit{secondary fluctuations}.  That is, the uncertainty in a quasiclassical variable exists only due to interacting with a quantum system; in the absence of such interaction, quasiclassical variables evolve in purely classical fashion.  This is in contrast to fully quantum variables, which exhibit primary fluctuations; those are fundamentally uncertain.

In this section, following Anderson's pioneering work, we provide an example of the evolution of (quasi)classical variables.  Quantum backreaction on classical variables is found in terms of a nonvanishing commutator between classical canonical conjugates.  We emphasize that here we only use the reduction property (Eq.~\eqref{eq:reduction}) and not a specific definition of the hybrid bracket.

It should be stressed that the following example only serves as an explicit illustration of the more general possibilities of hybrid dynamics.  The same logic applies to any Hamiltonian~\eqref{eq:hamiltonian}, even if more involved equations of motion result.  The example presented here demonstrates the basic concepts.

Consider a quantum-classical harmonic oscillator Hamiltonian
\begin{align}\label{eq:qc-ho}
    H = \frac{p_C^2}{2m_C} + \frac{p_Q^2}{2m_Q} + \frac{1}{2} k (x_C - x_Q)^2~.
\end{align}
The mass, position and momentum of the classical and quantum particles are given by $(m_C,x_C,p_C)$ and $(m_Q,x_Q,p_Q)$ respectively, and $k$ is the coupling strength.  Using the center-of-mass and relative separation coordinates
\begin{align}
\begin{split}
    X &= \frac{m_C \, x_C + m_Q \, x_Q}{m_C + m_Q}~,\quad
    P = p_C + p_Q \\
    x &= x_C - x_Q~,\qquad\qquad
    p = \frac{m_Q \, p_C - m_C \, p_Q}{m_C + m_Q}~,
\end{split}
\end{align}
and the total and reduced masses
\begin{align}
    M = m_C + m_Q~,\qquad\quad
    m = \frac{m_C \, m_Q}{m_C+m_Q}~,
\end{align}
the Hamiltonian can be written as
\begin{align}
H = \frac{P^2}{2M} + \frac{p^2}{2m} + \frac{1}{2} k x^2~.
\end{align}
Notice that $(X,P,x,p)$ are hybrid variables.  However, they are \textit{additive} hybrids that do not directly couple quantum and classical variables.

Formally, the new variables obey the canonical relations in terms of the hybrid bracket.  This can be shown using only the general reduction property~\eqref{eq:reduction} and bilinearity~\eqref{eq:lin}~:
\begin{align}
    \{\![X,P]\!\} &= \frac{1}{M} \{\![m_C x_C + m_Q x_Q, p_C + p_Q]\!\} \\
    &= \frac{m_C}{M} \{x_C, p_C\} + \frac{1}{i\hbar} \frac{m_Q}{M} [x_Q, p_Q] = 1~,
\end{align}
and similarly for the rest of the relations.  Using the hybrid fundamental relations, we can show that $\dot{X} = \{\![X,H]\!\}$ by explicit calculation
\begin{align}
    \dot{X} &= \frac{m_C}{M} \dot{x}_C + \frac{m_Q}{M} \dot{x}_Q \\
    &= \frac{m_C}{M} \{x_C,H\} + \frac{1}{i\hbar} \frac{m_Q}{M} [x_Q,H] \\
    &= \{\![X,H]\!\}~.
\end{align}
The reduction and bilinearity properties were again used to obtain the last equality.
    
Now the system can be solved using standard methods.  The center-of-mass position and momentum evolve according to
\begin{align}
    P = P(0)~, \qquad
    X = X(0) + \frac{P(0)}{M} t~.
\end{align}
The relative separation position and momentum equations of motion are
\begin{align}
    x &= x(0) \cos(\omega\,t) + \frac{p(0)}{m\omega} \sin(\omega\,t)~, \\
    p &= p(0) \cos(\omega\,t) - m\omega\, x(0) \sin(\omega\,t)~,
\end{align}
where $\omega = \sqrt{k/m}$~.  Finally, the equations of motion for the classical position and momentum are found to be 
\begin{align}\label{eq:xc}
    \begin{split}
    x_C =&~ \frac{1}{M} \big[m_C + m_Q \cos(\omega\,t) \big] x_C(0) \\
    &+ \frac{1}{m_CM\omega} \big[ m_C \omega\,t + m_Q \sin(\omega\,t) \big] p_C(0) \\
    &+ \frac{m_Q}{M} \big[ 1 - \cos(\omega\,t) \big] x_Q(0)\\
    &+ \frac{1}{M\omega} \big[ \omega\,t - \sin(\omega\,t) \big] p_Q(0)
    \end{split}
\end{align}
and
\begin{align}\label{eq:pc}
    \begin{split}
    p_C =&~ \frac{1}{M} \big[m_C + m_Q \cos(\omega\,t) \big] p_C(0) - m\omega \sin(\omega\,t) x_C(0) \\
    &+ \frac{m_C}{M} \big[ 1 - \cos(\omega\,t) \big] p_Q(0) + m\omega \sin(\omega\,t) x_Q(0)~.
    \end{split}
\end{align}
The equations of motion for $x_Q$ and $p_Q$ are identical and can be found by swapping subscripts $Q$ and $C$ in~\eqref{eq:xc} and~\eqref{eq:pc}.

\begin{figure}
\includegraphics[width=240px]{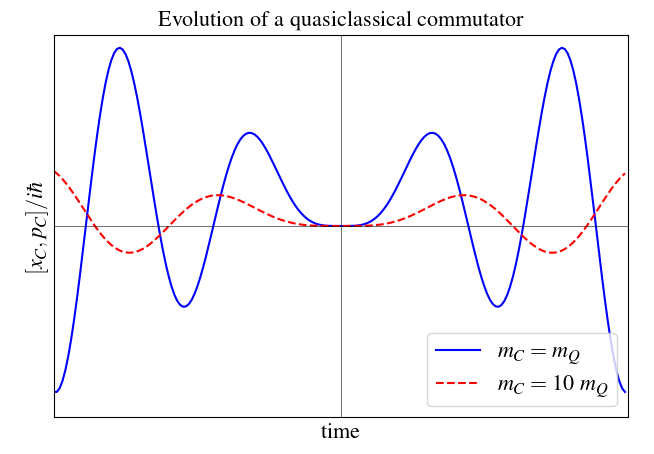}
\caption{Qualitative evolution of $[x_C,p_C]/i\hbar$ through Eq.~\eqref{eq:c-comm}, with $m_Q$ and $k$ set to unity.  At $t=0$, classical variables are known with certainty (vanishing commutator).  Before and after this moment of certainty, the commutator deviates from zero implying uncertain prediction/retrodiction.  As the classical mass $m_C$ becomes larger, the curve flattens approaching zero around the point of certainty at $t=0$.}
\label{fig:1}
\end{figure}

We see from the dependence of $x_C$ and $p_C$ on $x_Q(0)$ and $p_Q(0)$ that a form of backreaction is present on the classical variables.  While the commutator $[x_C,p_C]$ vanishes initially, it evolves to be non-zero due to interaction:
\begin{align}\label{eq:c-comm}
    \frac{1}{i\hbar}[x_C,p_C] = \frac{m}{M} \big( 2 - 2 \cos(\omega\,t) - \omega\,t \sin(\omega\,t) \big) ~.
\end{align}
If a nonvanishing commutator indicates uncertainty, then the above equation shows how quantum backreaction is manifested in a quantum-classical harmonic oscillator.  Fig.~\ref{fig:1} depicts the evolution of the commutator of $x_C$ and $p_C$.  Initially, $x_C(0)$ and $p_C(0)$ were known with complete certainty, perhaps through measurement.  Prediction or retrodiction at times other than $t=0$ is uncertain.  We see that as the classical mass $m_C$ becomes large compared to the quantum mass $m_Q$, the commutator $[x_C,p_C]$ approaches zero and classical behaviour dominates $x_C$ and $p_C$.

Further, non-classical behaviour is shown in the change of the value of the Poisson bracket.  While $\{x_C,p_C\}$ initially obeys the classical canonical relation, it evolves to deviate from unity
\begin{align}\label{eq:c-poisson}
\begin{split}
    \{x_C,p_C\} &= \frac{1}{M^2} \big(m_C + m_Q \cos(\omega\,t) \big)^2 \\
    &\quad+ \frac{m}{m_C M} \sin(\omega\,t) \big(m_C \omega\,t + m_Q \sin(\omega\,t) \big)~.
\end{split}
\end{align}
That deviation from classicality is, again, due to quantum backreaction.

The hybrid nature of $x_C$ and $p_C$ is now shown elegantly in terms of the hybrid bracket.  While they do not obey either the quantum or the classical canonical relations (Eqs.~\eqref{eq:c-comm} and~\eqref{eq:c-poisson}), they do obey the \textit{hybrid} fundamental canonical relation
\begin{align}
    \{\![x_C,p_C]\!\} = 1~.
\end{align}
Since $x_C$ and $p_C$ at $t \neq 0$ are additive hybrids, this can be shown explicitly using the reduction~\eqref{eq:reduction} and bilinearity~\eqref{eq:lin} properties applied to Eqs.~\eqref{eq:xc} and~\eqref{eq:pc} for $x_C$ and $p_C$.  Any hybrid bracket possessing these properties will produce the same results.  A similar calculation can be done for $x_Q$ and $p_Q$.

An important note is in order.  Thanks to the reduction property~\eqref{eq:reduction}, hybrid evolution of pure variables is consistent.  That is, if the hybrid bracket has one of its two arguments pure (or a sum of pure variables), it reduces to a consistent pure bracket.  This will be the case \textit{if only one hybrid variable is used in the problem}.  This covers a wide class of problems.  Usually the one interesting hybrid variable is the interaction Hamiltonian, while all other variables of interest are either pure classical or pure quantum such as positions, momenta, fields etc.  In the next section we will demonstrate the use of the full hybrid bracket if one desires to study the evolution of general hybrid variables.

\section{Evolution of hybrid variables}
\label{sec:4}

As demonstrated in the previous section, the evolution of pure variables and, by extension, additive hybrid variables is largely bracket-agnostic.  For nontrivial hybrids, however, a specific definition of the hybrid bracket is necessary.  Consider, for example, a hybrid variable $\eta$ that is a product of quantum and classical variables
\begin{align}
    \eta = \eta_C \eta_Q~.
\end{align}
On one hand, taking the time derivative of such a variable is unclear, since the Leibniz rule for derivation now runs into operator ordering ambiguities.  As seen in the previous section, classical variables will evolve into quasiclassical variables that do not commute with quantum ones.  Thus, while $\eta = \eta_C \eta_Q = \eta_Q \eta_C$, one does not know whether $\dot{\eta} = \dot{\eta}_C \eta_Q + \eta_C \dot{\eta}_Q$ or $\dot{\eta} = \eta_Q \dot{\eta}_C + \eta_C \dot{\eta}_Q$, for example.  On the other hand, the reduction property~\eqref{eq:reduction} is clearly insufficient to calculate the hybrid bracket when both variables are nontrivial hybrids so that $\{\![\eta,H]\!\}$ isn't defined either.  A consistent Leibniz rule for the hybrid bracket is needed.

The problem intensifies as various no-go theorems assert that the Leibniz rule and the Jacobi identity are not generally satisfied by hybrid brackets, making them inconsistent~\cite{salcedo_absence_1996,caro_impediments_1999,sahoo_mixing_2004,salcedo_statistical_2012,gil_canonical_2017}.  However, the authors have proposed a possible reinterpretation of these no-go theorems in light of the work presented in~\cite{amin_quantum-classical_2020}.  A general hybrid bracket is derived there, of the form
\begin{align}\label{eq:hybrid1}
    \{\![A,B]\!\} = \frac{1}{i\hbar} \big( A \circledast B - B \circledast A \big)~,
\end{align}
where $\circledast$ is the \textit{hybrid composition product}.  Now the consistency of the bracket is seen in light of this new composition product.  If $\circledast$ is associative, the bracket will obey the Jacobi identity, and the Leibniz rule will take the form
\begin{align}
    \{\![A, B \circledast C]\!\} = \{\![A, B]\!\} \circledast C + B \circledast \{\![A, C]\!\}~.
\end{align}

Now that the problem is cast in definite terms, we propose a possible circumvention.  If we restrict hybrid variables to be only those forming an associative subalgebra with the hybrid composition product $\circledast$, then the dynamics of these variables is consistent.  This implies that only certain quantum-classical interactions are allowed, as dictated by the consistency of the framework.

Since associativity with the $\circledast$-product is the central condition, a concrete definition of $\circledast$ is needed.  In~\cite{amin_quantum-classical_2020}, it is shown that a general hybrid bracket is derived through the application of a partial classical limit to a full quantum theory.  A quantum system is subdivided into $Q$(uantum) and $C$(lassical) sectors.  Then a classical limit is applied on the $C$ sector using the phase space formulation of quantum mechanics.  The form of $\circledast$ depends on the choice of the quantization scheme (e.g., the ordering recipe) on the $C$ sector.  

The $\circledast$-product~\cite{amin_quantum-classical_2020}
\begin{align}\label{eq:ast}
    A \circledast B = AB + \frac{i\hbar}{2} \big[ \{A, B\} + \sigma(A, B) \big]
\end{align}
acts only on the classical part of hybrid variables.  The symmetric binary operation $\sigma$ reflects the quantization scheme on the $C$ sector prior to taking the classical limit.  Using~\eqref{eq:ast} in~\eqref{eq:hybrid1}, we obtain the bracket in a more familiar form
\begin{align}\label{eq:hybrid2}
\begin{split}
    \{\![A,B]\!\} = \frac{1}{i\hbar} [A,B] &+ \frac{1}{2} \big[ \{A,B\} - \{B,A\} \big] \\
    &+ \frac{1}{2} \big[ \sigma(A,B) - \sigma(B,A) \big]~.
\end{split}
\end{align}
For $\sigma = 0$, the bracket reduces to that proposed by Aleksandrov~\cite{aleksandrov_statistical_1981}, Gerasimenko~\cite{gerasimenko_dynamical_1982}, and Boucher and Traschen~\cite{boucher_semiclassical_1988}.

A general formula for $\sigma$ resulting from familiar quantization schemes on the $C$ sector is given by
\begin{align}\label{eq:sigma}
\begin{split}
    \sigma({A},{B}) &= a ~ \frac{\partial A}{\partial x_C}\,\frac{\partial B}{\partial x_C}
    + b ~ \frac{\partial A}{\partial p_C}\,\frac{\partial B}{\partial p_C} \\
    &\qquad + c \left( \frac{\partial A}{\partial x_C}\,\frac{\partial B}{\partial p_C} + \frac{\partial A}{\partial p_C}\,\frac{\partial B}{\partial x_C} \right)~,
\end{split}
\end{align}
where the constants $(a,b,c)$ reflect the choice of quantization.  For example, $(0,0,0)$ reflects the Weyl ordering associated with the Wigner phase space distribution, while $(1,1,0)$ reflects a certain parameterization of the anti-normal ordering associated with the Husimi distribution.

Now a hybrid equation of motion for hybrid variables in the Heisenberg picture can be defined as
\begin{align}
    \frac{dA}{dt} = \{\![A,H]\!\} + \frac{\partial A}{\partial t}~.
\end{align}
Once a choice for $\sigma$ (or the quantization scheme prior to the classical limit) is specified, dynamical evolution is unambiguously defined.  Different $\sigma$'s will produce different evolutions.

As an illustrative example, take
\begin{align}
    \eta = p_C\, p_Q
\end{align}
to be a hybrid variable of interest.  Notice that the time derivative of the classical part of $\eta$ does not commute with quantum variables.  Given the harmonic oscillator Hamiltonian~\eqref{eq:qc-ho}, we have
\begin{align}
    \dot{\eta}_C = \dot{p}_C = \{p_C,H\} = -k (x_C - x_Q)~,
\end{align}
and thus the ordering of $\frac{d}{dt} \left( \eta_Q \eta_C \right)$ is ambiguous.  To see the problem, let us naively calculate $\dot{\eta}$ in two different orderings
\begin{subequations}\label{eq:naive}
\begin{align}
\begin{split}
    \left( \dot{\eta} \right)_1 &= \dot{p}_C\, p_Q + p_C\, \dot{p}_Q \\
    &= k \left( x_Q \, p_Q - x_C \, p_Q - x_Q \, p_C + x_C \, p_C \right)~,
\end{split}
\end{align}
\begin{align}
\begin{split}
    \left( \dot{\eta} \right)_1 &= \dot{p}_C\, p_Q + p_C\, \dot{p}_Q \\
    &= k \left( p_Q \, x_Q - x_C \, p_Q - x_Q \, p_C + x_C \, p_C \right)~,
\end{split}
\end{align}
\end{subequations}
It could be suggested to symmetrize the product of $x_Q$ and $p_Q$, and thus get rid of the ambiguity.  However, such symmetrization would be an ad hoc choice of operator ordering on the $Q$ sector.  A consistent hybrid bracket is necessary.

In terms of the hybrid bracket~\eqref{eq:hybrid2}, time evolution of $\eta$ is given by
\begin{align}\label{eq:eta-dot}
\begin{split}
    \dot{\eta} &= \{\![\eta,H]\!\} \\
    &= k \left( \frac{x_Q \, p_Q + p_Q \, x_Q}{2} - x_C \, p_Q - x_Q \, p_C + x_C \, p_C \right. \\
    &\qquad \qquad \qquad \qquad \qquad \qquad  + \left. \frac{i\hbar}{2} \sigma(x_C,p_C) \right)~.
\end{split}
\end{align}
Compare~\eqref{eq:eta-dot} to~\eqref{eq:naive}.  While a symmetrization of~\eqref{eq:naive} gets rid of the ordering ambiguity (in an ad-hoc manner), it completely misses out on the $\sigma$ term present in~\eqref{eq:eta-dot}.  To reiterate, $\sigma$ is connected the operator ordering on the $C$ sector, not the $Q$ sector, before taking the classical limit.

As promised, different forms of $\sigma$ will have different evolutions.  In the simple example of $\eta=p_C\,p_Q$, using the definition~\eqref{eq:sigma} of $\sigma$, $\dot{\eta}$ becomes
\begin{align}\label{eq:eta-dot-c}
    k \left( \frac{x_Q \, p_Q + p_Q \, x_Q}{2} - x_C \, p_Q - x_Q \, p_C + x_C \, p_C + \frac{i\hbar}{2}c \right)~.
\end{align}
Here, $c$ is a constant determined by the original quantization scheme on the $C$ sector before the classical limit, as discussed below Eq.~\eqref{eq:sigma}.  Finally, the time dependence of $\eta$ can be found by plugging in the solutions for $(x_C,p_C,x_Q,p_Q)$.  The classical variables $x_C$ and $p_C$ have been explicitly calculated in Eqs.~\eqref{eq:xc} and~\eqref{eq:pc}.  The equations for $(x_Q,p_Q)$ can be found similarly.

The argument presented in this section is not specific to the harmonic oscillator, of course.  The procedure is general and applicable to any Hamiltonian and hybrid variable provided they belong to an associative subalgebra with the hybrid composition product $\circledast$.

\section{Conclusion}
\label{sec:conc}

The example of the quantum-classical harmonic oscillator presented here exhibits basic features of hybrid systems.  The treatment used is applicable to general Hamiltonians.

Quantum backreaction in interacting quantum-classical systems can, in principle, be described through the framework of Hamiltonian hybrid dynamics.  The example of the quantum-classical harmonic oscillator studied here demonstrates this backreaction: classical variables grow to be uncertain by interacting with quantum ones.  The uncertainty in classical variables is manifested in the nonvanishing commutator of classical variables.  

Along with Anderson~\cite{anderson_quantum_1995}, we note that despite the fuzziness of classical variables through interaction, they are still classical in the sense that their uncertainty is entirely due to the effect on them of truly quantum variables.  That effect is depicted in Fig.~\ref{fig:1}: a commutator of classical variables vanishes initially, but evolves to be nonzero at later times.  The figure also illustrates that the more massive the classical system is, the longer it remains certain.

An important property made plain by the analysis presented here is that the details of a hybrid bracket are irrelevant for the dynamics of pure variables.  All brackets obeying the reduction and bilinearity properties (Eqs.~\eqref{eq:reduction} and~\eqref{eq:lin}) are equivalent as far as pure variables are concerned.  This covers a wide class of problems where the only nontrivial hybrid variable is the interaction Hamiltonian.

Dynamics of hybrid variables requires the use of a fully-defined hybrid bracket.  Without such a bracket, we run into ordering ambiguities and miss extra effects resulting from the hybrid nature of the system.  As shown in~\cite{amin_quantum-classical_2020}, these extra effects arise from the specific quantization scheme (and its associated phase space distribution) used in deriving the bracket.

\section*{Acknowledgements}
    This research was funded partly by an NSERC Discovery Grant.

\newpage
\footnotesize \bibliography{refs}   

\begin{thebibliography}{10}

\bibitem{maddox_classical_1995}
John Maddox.
\newblock Classical and quantum physics mix.
\newblock {\em Nature}, 373, 1995.

\bibitem{aleksandrov_statistical_1981}
I.~V. Aleksandrov.
\newblock The {Statistical} {Dynamics} of a {System} {Consisting} of a
  {Classical} and a {Quantum} {Subsystem}.
\newblock {\em Zeitschrift für Naturforschung A}, 36, 1981.

\bibitem{gerasimenko_dynamical_1982}
V.~I. Gerasimenko.
\newblock Dynamical equations of quantum-classical systems.
\newblock {\em Theor Math Phys}, 50, 1982.

\bibitem{boucher_semiclassical_1988}
Wayne Boucher and Jennie Traschen.
\newblock Semiclassical physics and quantum fluctuations.
\newblock {\em Phys. Rev. D}, 37, 1988.

\bibitem{anderson_quantum_1995}
Arlen Anderson.
\newblock Quantum {Backreaction} on "{Classical}" {Variables}.
\newblock {\em Phys. Rev. Lett.}, 74, 1995.

\bibitem{prezhdo_mixing_1997}
Oleg~V. Prezhdo and Vladimir~V. Kisil.
\newblock Mixing quantum and classical mechanics.
\newblock {\em Phys. Rev. A}, 56, 1997.

\bibitem{prezhdo_quantum-classical_2006}
Oleg~V. Prezhdo.
\newblock A quantum-classical bracket that satisfies the {Jacobi} identity.
\newblock {\em J. Chem. Phys.}, 124, 2006.

\bibitem{elze_linear_2012}
Hans-Thomas Elze.
\newblock Linear dynamics of quantum-classical hybrids.
\newblock {\em Phys. Rev. A}, 85, 2012.

\bibitem{bondar_koopman_2019}
Denys~I. Bondar, François Gay-Balmaz, and Cesare Tronci.
\newblock Koopman wavefunctions and classical–quantum correlation dynamics.
\newblock {\em Proc. R. Soc. A.}, 475(2229):20180879, September 2019.

\bibitem{barcelo_hybrid_2012}
Carlos Barceló, Raúl Carballo-Rubio, Luis~J. Garay, and Ricardo
  Gómez-Escalante.
\newblock Hybrid classical-quantum formulations ask for hybrid notions.
\newblock {\em Phys. Rev. A}, 86, 2012.

\bibitem{elze_four_2012}
Hans-Thomas Elze.
\newblock Four questions for quantum-classical hybrid theory.
\newblock {\em J. Phys.: Conf. Ser.}, 361, 2012.

\bibitem{elze_quantum-classical_2013}
Hans-Thomas Elze.
\newblock Quantum-classical hybrid dynamics - a summary.
\newblock {\em J. Phys.: Conf. Ser.}, 442, 2013.

\bibitem{salcedo_absence_1996}
L.~L. Salcedo.
\newblock Absence of classical and quantum mixing.
\newblock {\em Phys. Rev. A}, 54, 1996.

\bibitem{caro_impediments_1999}
J.~Caro and L.~L. Salcedo.
\newblock Impediments to mixing classical and quantum dynamics.
\newblock {\em Phys. Rev. A}, 60, 1999.

\bibitem{sahoo_mixing_2004}
Debendranath Sahoo.
\newblock Mixing quantum and classical mechanics and uniqueness of {Planck}'s
  constant.
\newblock {\em J. Phys. A: Math. Gen.}, 37, 2004.

\bibitem{salcedo_statistical_2012}
L.~L. Salcedo.
\newblock Statistical consistency of quantum-classical hybrids.
\newblock {\em Phys. Rev. A}, 85, 2012.

\bibitem{gil_canonical_2017}
V.~Gil and L.~L. Salcedo.
\newblock Canonical bracket in quantum-classical hybrid systems.
\newblock {\em Phys. Rev. A}, 95, 2017.

\bibitem{amin_quantum-classical_2020}
Mustafa Amin and Mark~A. Walton.
\newblock Quantum-{Classical} {Dynamical} {Brackets}.
\newblock 2020.
\newblock arXiv:2009.09573 [quant-ph].

\end{thebibliography}

\end{document}